\documentclass[
aps,prb,
amsmath,amssymb,
superscriptaddress,
longbibliography,
reprint]{revtex4-1}

\usepackage{graphicx}

\begin{document}

\title{Background force compensation in dynamic atomic force microscopy}

\author{Riccardo Borgani}
\email{borgani@kth.se}
\affiliation{Nanostructure Physics, KTH Royal Institute of Technology, Stockholm, Sweden}

\author{Per-Anders Thor\'en}
\affiliation{Nanostructure Physics, KTH Royal Institute of Technology, Stockholm, Sweden}

\author{Daniel Forchheimer}
\affiliation{Nanostructure Physics, KTH Royal Institute of Technology, Stockholm, Sweden}

\author{Illia Dobryden}
\affiliation{Surface and Corrosion Science, KTH Royal Institute of Technology, Stockholm, Sweden}

\author{Si Mohamed Sah}
\affiliation{Nanostructure Physics, KTH Royal Institute of Technology, Stockholm, Sweden}

\author{Per Martin Claesson}
\affiliation{Surface and Corrosion Science, KTH Royal Institute of Technology, Stockholm, Sweden}

\author{David B. Haviland}
\affiliation{Nanostructure Physics, KTH Royal Institute of Technology, Stockholm, Sweden}

\date{December 16, 2016}

\begin{abstract}
Background forces are linear long range interactions of the cantilever body with its surroundings that must be compensated for in order to reveal tip-surface force, the quantity of interest for determining material properties in atomic force microscopy.
We provide a mathematical derivation of a method to compensate for background forces, apply it to experimental data, and discuss how to include background forces in simulation.
Our method, based on linear response theory in the frequency domain, provides a general way of measuring and compensating for any background force and it can be readily applied to different force reconstruction methods in dynamic AFM.
\end{abstract}

\maketitle

\section{Introduction}

Accurate and reproducible measurement of material properties at the nanoscale is the main goal of dynamic atomic force microscopy (AFM).
Extraction of material properties from the measurable quantities in dynamic AFM requires a deep understanding of both the tip-surface interaction and the dynamics of the AFM cantilever when it is close to the sample surface.
We propose a method that uses Fourier analysis to measure and compensate for background forces, which are long range and not local to the AFM tip.
These interactions produce artifacts in the measurement of tip-surface force, leading to overestimation of its attractive and dissipative components.

\begin{figure}
  \includegraphics[width=\columnwidth]{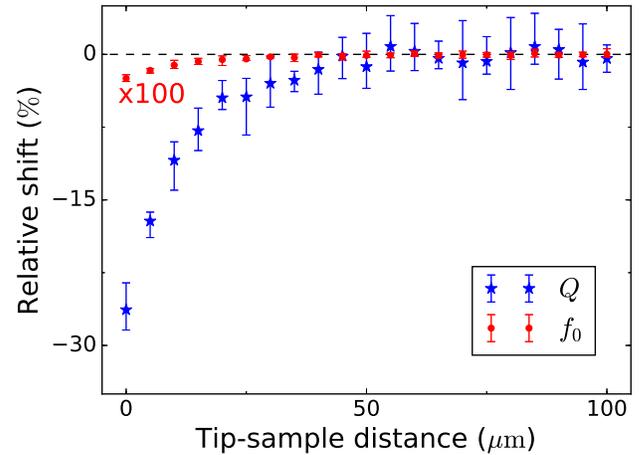}
  \caption{Relative change in quality factor $Q$ and resonance frequency $f_0$ as a function of tip-sample distance over a homogeneous PDMS surface.
  The values are obtained by fitting the thermal noise power spectral density, and plotted as the shift relative to a tip-sample distance of $1~\mathrm{mm}$.
  The fitted $Q$ drops by as much as $30\%$, whereas the decrease in $f_0$ is three orders of magnitude smaller (plotted values of $f_0$ are multiplied by~$100$).
  The cantilever was a MikroMasch HQ:NSC15/AlBS.}
  \label{fig:Qf0}
\end{figure}

Background forces are observed when measuring the quality factor of a cantilever resonance which decreases when the tip-sample distance becomes comparable to the cantilever width, dropping as much as~$30\%$ at the scanning position (Fig.~\ref{fig:Qf0}).
This phenomenon has been attributed to an additional squeeze-film damping force\cite{Zhang2009,Honig2010,Sader2016}, arising when the fluid surrounding the cantilever is squeezed between the cantilever body and the sample surface.
We also observe a slight decrease in the resonance frequency $f_0$, due to increased hydrodynamic load.

Other long range forces appearing at tip-sample distances of a few micrometers have been attributed to electrostatic contributions\cite{Law2002}.
We have also observed in different commercial AFM systems, a dependence of the cantilever's acoustic excitation on the extension of the scanning z-piezo, resulting in a change of the drive force which can be mistaken as a long range interaction.
The coupling between the cantilever and the acoustic actuator could in principle also be affected by the increased hydrodynamic load, causing the effective drive power to change as a function of the tip-sample distance.

Whatever their origin, be it hydrodynamic, electrostatic, or AFM design, these effects influence force reconstruction in dynamic AFM.
Previous attempts at compensating for them have used an effective resonance frequency and quality factor for the cantilever\cite{Jesse2007,Killgore2011,Tung2014}, but a general and accurate description is still lacking.
Here we describe how to compensate for these effects by treating them as background forces, assuming that they have the following properties: they are linear as shown by a lack of intermodulation distortion\cite{Platz2008}; they act over a long range, comparable to the cantilever width; and they do not depend on the $xy$ tip position over the sample surface, as they originate from the cantilever body rather than being local to the tip.

We can easily compensate for any such background force to reveal the true tip-surface interaction, using linear response theory in the frequency domain.
Due to its generality and ease of implementation, we expect our method to be readily applied to a variety of force reconstruction methods essential for AFM researchers.

\section{\label{sec:chi_theory} Generic linear model}

\begin{figure}
  \includegraphics[width=\columnwidth]{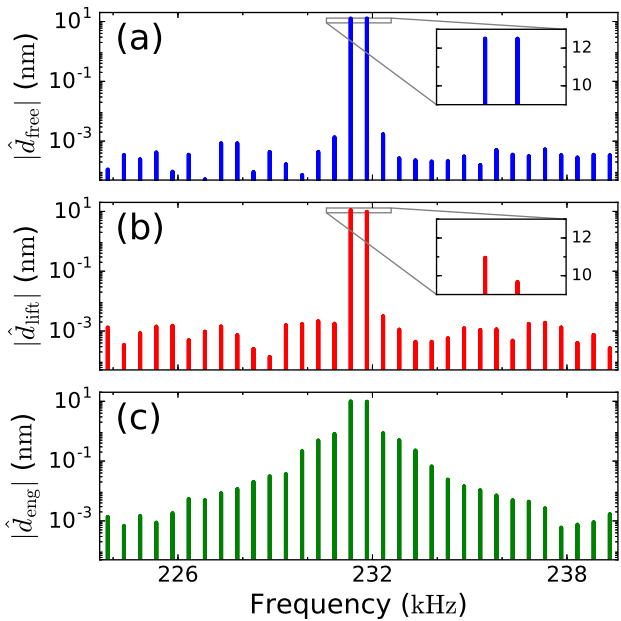}
  \caption{Discrete frequency spectra of cantilever motion measured near resonance using a 2-tone drive.
  Phase is also measured at each frequency but only amplitude is shown.
  (a) At the free position far away from the sample surface.
  (b) At the lift position closer to the surface.
  (c) At the engaged position on a polystyrene surface.
  In (a) and (b) linear forces act on the cantilever and only noise is measured at the undriven frequencies.
  In (c) the nonlinear tip-surface force gives rise to intermodulation with strong response at undriven frequencies.}
  \label{fig:spectra}
\end{figure}

When an AFM cantilever is far away from the sample surface, its fundamental flexural mode is well-modeled as a linear system.
The frequency dependent linear response function $\hat{\chi}(\omega)$ relates the frequency components of any force $\hat{F}(\omega)$ to the resulting frequency components of the cantilever deflection $\hat{d}(\omega)$:
\begin{equation}
  \hat{d}(\omega) = \hat{\chi}(\omega) \hat{F}(\omega)
  .
\end{equation}
In our notation $\hat{d}(\omega)$ denotes a complex valued function of the real variable $\omega$, the frequency.
In particular, $\hat{d}(\omega)$ is the Fourier transform of $d(t)$.
In the following, we will drop the explicit $\omega$ dependence for the sake of compact notation.

When the frequency components of $\hat{d}$ are concentrated around the cantilever resonance frequency, $\hat{\chi}$ can be well-modeled as a dampened simple harmonic oscillator:
\begin{equation}
  \hat{\chi}^{-1} = k \left( 1 - \frac{\omega^2}{\omega_0^2} + \mathrm{i}\frac{\omega}{Q\omega_0} \right)
  \label{eq:chi_cant}
  .
\end{equation}
The parameters $k$, $\omega_0$ and $Q$ are the mode stiffness, resonance frequency and quality factor, respectively.
These parameters, together with the optical lever responsivity, can be obtained by a non-invasive calibration procedure traceable to the measurement of the thermal fluctuations of the cantilever deflection when it is far from the surface\cite{Sader2012,Higgins2006}.

The drive force with multiple frequency components $\hat{F}_\mathrm{D}$ is applied to the cantilever by means of a shaker piezo in the case of acoustic actuation, or a pulsed laser beam in the case of photo-thermal excitation.
Regardless of the means of excitation, the drive force is determined by measuring the cantilever motion far away from the surface (larger than $100~\mathrm{\mu m}$) at what we call the ``free'' position (Fig.~\ref{fig:spectra}(a)).
Thus, we extract the driving force from a measurement of the cantilever free motion $\hat{d}_\mathrm{free}$ and the calibrated linear response function $\hat{\chi}$
\begin{equation}
  \hat{F}_\mathrm{D} = \hat{\chi}^{-1} \hat{d}_\mathrm{free}
  \label{eq:f_drive}
  .
\end{equation}

As the AFM probe approaches the sample surface, background forces begin to affect the cantilever body when its separation from the surface becomes comparable to its width, as shown in Fig.~\ref{fig:Qf0}.
Background forces change the cantilever motion (compare insets of Fig.~\ref{fig:spectra}(a) and Fig.~\ref{fig:spectra}(b)), but they are clearly linear, as seen from the lack of intermodulation\cite{Platz2008} in the spectrum of Fig.~\ref{fig:spectra}(b).

When the AFM tip starts interacting with the surface at what we call the ``engaged'' position (Fig.~\ref{fig:spectra}(c)), the measured motion is affected by all the forces at play
\begin{equation}
  \hat{d}_\mathrm{eng} = \hat{\chi} \left( \hat{F}_\mathrm{D}
                         + \hat{F}_\mathrm{BG}
                         + \hat{F}_\mathrm{TS} \right)
  ,
  \label{eq:all_forces}
\end{equation}
where $\hat{F}_\mathrm{TS}$ is the nonlinear tip-surface force, carrying all the information about the material properties, and $\hat{F}_\mathrm{BG}$ are the background forces.
We can use Eq.~(\ref{eq:f_drive}) to account for the drive force, but in oder to solve for the tip-surface force we must eliminate the background forces.

Note that, while the components of the drive force $\hat{F}_\mathrm{D}$ can be treated as constant, the components of the background forces $\hat{F}_\mathrm{BG}$ depend on the motion $\hat{d}$ which changes from pixel to pixel.
For example, a squeeze-film damping force will depend on the velocity of the cantilever.

Motivated by experimental observation (lack of intermodulation distortion), we treat the problem of a general linear background force without regard to its particular origin, by expressing it in terms of a linear response function $\hat{\chi}_\mathrm{BG}$
\begin{equation}
  \hat{F}_\mathrm{BG} = \hat{\chi}_\mathrm{BG}^{-1} \hat{d}
  \label{eq:def_chi_bg}
  .
\end{equation}
Equation~(\ref{eq:def_chi_bg}) allows for the calculation of $\hat{F}_\mathrm{BG}$ for any motion $\hat{d}$.
Our treatment assumes that there exists a linear differential equation of the cantilever deflection which describes the background forces.
This equation can in principle be very complicated, \textit{e.g.} involve fractional derivatives and have many parameters, but we assume that it does not change as the probe scans over the sample, consistent with the idea that the background forces act on the body of the cantilever.

Lifting the probe slightly away from the surface, we find that the short ranged $\hat{F}_\mathrm{TS}$ goes to zero with an abrupt drop in intermodulation, while the long ranged $\hat{F}_\mathrm{BG}$ is barely affected.
We define the ``lift'' position as the closest distance for which the forces acting on the cantilever are linear (see Sect.~\ref{sec:algorithm}).

At the lift position (Fig.~\ref{fig:spectra}(b)), the total force is given by the drive force $\hat{F}_\mathrm{D}$ and the linear background forces $\hat{F}_\mathrm{BG}$ only.
The lift motion is therefore
\begin{equation}
  \hat{d}_\mathrm{lift} = \hat{\chi} \hat{F}_\mathrm{D}
                          + \hat{\chi} \hat{F}_\mathrm{BG}
  .
\end{equation}
Solving for $\hat{F}_\mathrm{BG}$ gives
\begin{equation}
  \hat{F}_\mathrm{BG} = \hat{\chi}^{-1} \hat{d}_\mathrm{lift} - \hat{F}_\mathrm{D}
  \label{eq:f_bg_lift}
  .
\end{equation}

Combining Eq.~(\ref{eq:f_drive}), Eq.~(\ref{eq:def_chi_bg}), and Eq.~(\ref{eq:f_bg_lift}), we determine $\hat{\chi}_\mathrm{BG}$ from the measured $\hat{d}_\mathrm{free}$ and $\hat{d}_\mathrm{lift}$
\begin{equation}
  \hat{\chi}_\mathrm{BG}^{-1} = \hat{\chi}^{-1} \frac{\hat{d}_\mathrm{lift} - \hat{d}_\mathrm{free}}{\hat{d}_\mathrm{lift}}
  \label{eq:chi_bg}
  .
\end{equation}
Going back to the engaged position, we can now compensate for the background forces in Eq.~(\ref{eq:all_forces}) using $\hat{\chi}_\mathrm{BG}$ from Eq.~(\ref{eq:chi_bg}). Thus we obtain the tip-surface force:
\begin{equation}
  \hat{F}_\mathrm{TS} = \hat{\chi}^{-1} \hat{d}_\mathrm{eng} - \hat{F}_\mathrm{D} - \hat{\chi}_\mathrm{BG}^{-1} \hat{d}_\mathrm{eng}
  \label{eq:correction}
  .
\end{equation}
Equation~(\ref{eq:correction}) allows for the compensation of the background forces, and thus the calculation of the tip-surface force at every pixel of an AFM image, provided the knowledge of $\hat{F}_\mathrm{D}$ and of $\hat{\chi}_\mathrm{BG}$, both being constant during the AFM scan.

\section{\label{sec:algorithm} Defining the lift position}

\begin{figure}
    \includegraphics[width=\columnwidth]{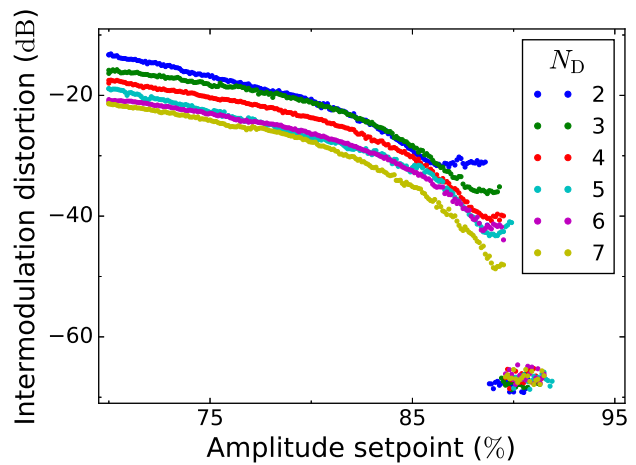}
    \caption{Intermodulation distortion ($\mathit{IMD}$) for different number of drive tones $N_\mathrm{D}$ is measured as function of the AFM feedback amplitude setpoint.
    As the AFM probe moves away from the surface (increasing setpoint), $\mathit{IMD}$ gradually decreases until the probe suddenly breaks free from the tip-surface interaction $F_\mathrm{TS}$ and a sharp drop in $\mathit{IMD}$ is observed.
    The lift motion is measured at this drop.
    For all data shown, the total free amplitude was $50~\mathrm{nm}$.
    }
    \label{fig:imd}
\end{figure}

To accurately determine the linear response function of the background forces $\hat{\chi}_\mathrm{BG}$, we need to measure the cantilever motion $\hat{d}_\mathrm{lift}$ as close to the surface as possible without tip-surface interaction.
We use the notion that the tip-surface force is strongly nonlinear, while the background forces are linear, as evidenced by the measurements of Fig.~\ref{fig:spectra} and Fig.~\ref{fig:imd}.

We apply a multifrequency drive with a number of discrete components $N_\mathrm{D}$ at the set of frequencies $\{\omega_{\mathrm{D}k}\}$ with $N_\mathrm{D} \ge 2$, \textit{i.e.} $\hat{F}_\mathrm{D}(\omega)$ is non-zero for $\omega \in \{\omega_{\mathrm{D}k}\}$.
When the linear forces $\hat{F}_\mathrm{D}$ and $\hat{F}_\mathrm{BG}$ act on the cantilever, the response $\hat{d}(\omega)$ will be non-zero only at $\omega \in \{\omega_{\mathrm{D}k}\}$ (see Fig.~\ref{fig:spectra}(a,b)).
On the other hand, when the cantilever is experiencing the nonlinear $\hat{F}_\mathrm{TS}$, response will arise at intermodulation product frequencies $\omega_\mathrm{IMP}$ given by integer linear combinations of the drive frequencies:

\begin{equation}
  \omega_\mathrm{IMP} = \sum_{k=1}^{N_\mathrm{D}} n_k \omega_{\mathrm{D}k},~~~~~n_k \in \mathbb{Z}
  ,
\end{equation}
where $n_k$ is an integer (see Fig.~\ref{fig:spectra}(c)).

We introduce the intermodulation distortion $\mathit{IMD}$, as the ratio of the power at undriven frequencies to the power at driven frequencies:

\begin{equation}
  \mathit{IMD} = \frac{\sum_{\omega_\mathrm{IMP} \notin \{\omega_{\mathrm{D}k}\}} \left|\hat{d}(\omega_\mathrm{IMP})\right|^2}{\sum_{\omega_\mathrm{IMP} \in \{\omega_{\mathrm{D}k}\}} \left|\hat{d}(\omega_\mathrm{IMP})\right|^2}
  .
\end{equation}

In principle, we want to measure $\hat{d}_\mathrm{lift}$ at the minimum distance from the surface such that $\mathit{IMD}=0$.
In practice, however, we will always measure some non-zero noise power.
We therefore choose a threshold (typically $3~\mathrm{dB}$) and measure $\hat{d}_\mathrm{lift}$ at the minimum distance from the surface such that $\mathit{IMD}_\mathrm{lift}<\mathit{IMD}_\mathrm{free} + \mathit{threshold}$.

Figure~\ref{fig:imd} shows the intermodulation distortion as a function of amplitude setpoint measured for drive schemes with different number of drive tones $N_\mathrm{D}$.
As the AFM feedback setpoint increases, a characteristic behavior is visible showing a gradual decrease of $\mathit{IMD}$ with increasing setpoint, due to the decrease of the non-linear tip-surface interaction $\hat{F}_\mathrm{TS}$.
When the setpoint reaches a value of about $90\%$, a sharp drop in $\mathit{IMD}$ is observed, indicating a transition from an overall nonlinear force to an overall linear force.
This sharp transition allows for unambiguous measurement of $\hat{d}_\mathrm{lift}$, from which we calculate the linear response function of the background forces $\hat{\chi}_\mathrm{BG}$.

\section{Extrapolation to undriven frequencies}

As discussed in Sect.~\ref{sec:algorithm}, $\hat{d}_\mathrm{lift}$ and $\hat{d}_\mathrm{free}$ will be non-zero only at the drive frequencies.
Calculating the linear response function of the background forces from Eq.~(\ref{eq:chi_bg}) will therefore yield $\hat{\chi}_\mathrm{BG}(\omega)$ only at the drive frequencies $\omega \in \{\omega_{\mathrm{D}k}\}$.
On the other hand, to apply the compensation to the measured data with Eq.~(\ref{eq:correction}) we require the knowledge of $\hat{\chi}_\mathrm{BG}(\omega)$ at all the frequencies in the spectrum of engaged motion $\hat{d}_\mathrm{eng}$.

To overcome this issue we use the notion of narrow band measurement on a resonant system.
Due to the high Q resonance in the cantilever linear response function, the motion will be concentrated at frequencies close to the resonance frequency $\omega_0$, within a narrow band~$\Omega$
\begin{equation}
  \Omega \approx N_\mathrm{IMP}\frac{\omega_0}{Q} \ll \omega_0
  ,
\end{equation}
where $N_\mathrm{IMP}$ is the number of measured frequencies (typically 32) and $\omega_0/Q$ is typically chosen as the measurement bandwidth.
We perform a polynomial expansion of the complex function $\hat{\chi}_\mathrm{BG}(\omega)$ in this narrow band:
\begin{equation}
  \hat{\chi}_\mathrm{BG}^{-1}(\omega) \approx \sum_{k = 0}^{M} (a_k + \mathrm{i}b_k) (\omega - \omega_0)^k
  \label{eq:polyfit}
  ,
\end{equation}
where $\mathrm{i}$ is the imaginary constant and $\{a_k\}$ and $\{b_k\}$ are sets of real coefficients to be determined.
A drive force with $N_\mathrm{D}$ frequency components allows for the determination of up to $2N_\mathrm{D}$ coefficients, corresponding to two polynomial fits of degree $M = N_\mathrm{D}-1$ of the real and imaginary parts of $\hat{\chi}_\mathrm{BG}$.
It is possible to perform a low degree fit with a high number of drives, $M < N_\mathrm{D}-1$, in which case the coefficients are obtained with a least square optimization method.

While a higher order fit could in principle describe a more complex $\hat{\chi}_\mathrm{BG}$, we find that a linear approximation of each of the two quadratures (4 coefficients, requiring 2 or more drive frequencies) is sufficient to describe the background forces.
A higher order fit is not always numerically stable, and it can introduce artifacts in the compensated data.

Equation~(\ref{eq:polyfit}) is quite general, allowing for a good approximation to any type of linear background force.
A special case of Eq.~(\ref{eq:polyfit}) is a polynomial with only two coefficients of the form
\begin{equation}
  \hat{\chi}_\mathrm{BG}^{-1}(\omega) \approx k (a \omega^2 + \mathrm{i} b \omega)
  \label{eq:polyfit_ho}
  ,
\end{equation}
where $k$ is the mode stiffness and $a$ and $b$ are fit parameters.
In this case it can be shown that the background forces result in an effective cantilever with a renormalized linear response function $\hat{\chi}^\prime$ of the form of Eq.~(\ref{eq:chi_cant}), where the resonance frequency and quality factor are given by

\begin{align}
  \omega_0^\prime &= \omega_0 \frac{1}{\sqrt{1-a\omega_0^2}}
  \label{eq:w0_eff}
  ,\\
  Q^\prime &= Q \frac{\sqrt{1-a\omega_0^2}}{1+bQ\omega_0}
  \label{eq:Q_eff}
  .
\end{align}

This special case is often assumed when analyzing forces in dynamic AFM\cite{Jesse2007,Killgore2011,Tung2014}.

\section{Experimental results}

\begin{figure*}
    \includegraphics[width=2\columnwidth]{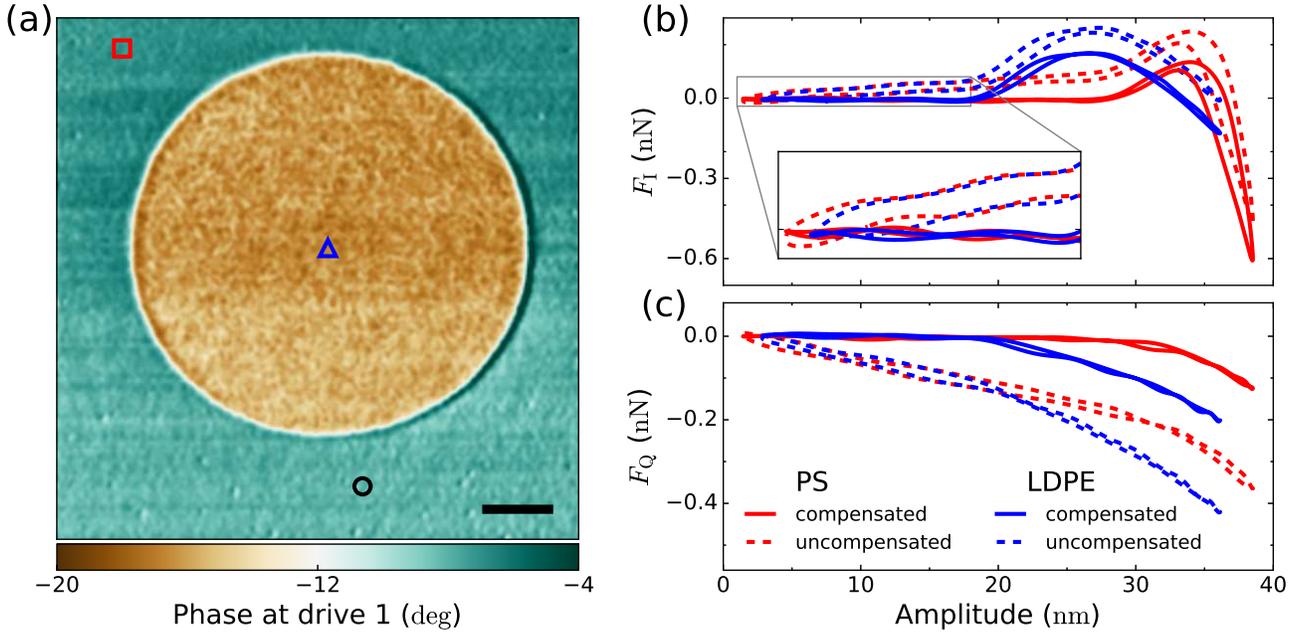}
    \caption{(a) Phase image at the first drive frequency on a blend of polystyrene (PS) and low-density polyethylene (LDPE). The blue triangle and the red square mark the pixels for which the engaged spectrum is analyzed. The black circle marks the location where the lift motion was measured using the method described in Sect.~\ref{sec:algorithm}. The scale bar is $200~\mathrm{nm}.$ (b, c) Dynamic force quadratures on PS (red) and on LDPE (blue) at the pixels marked in the corresponding color. Dashed lines show uncompensated measurements and solid lines show compensation for background forces.}
    \label{fig:PS_LDPE}
\end{figure*}

\begin{figure*}
    \includegraphics[width=2\columnwidth]{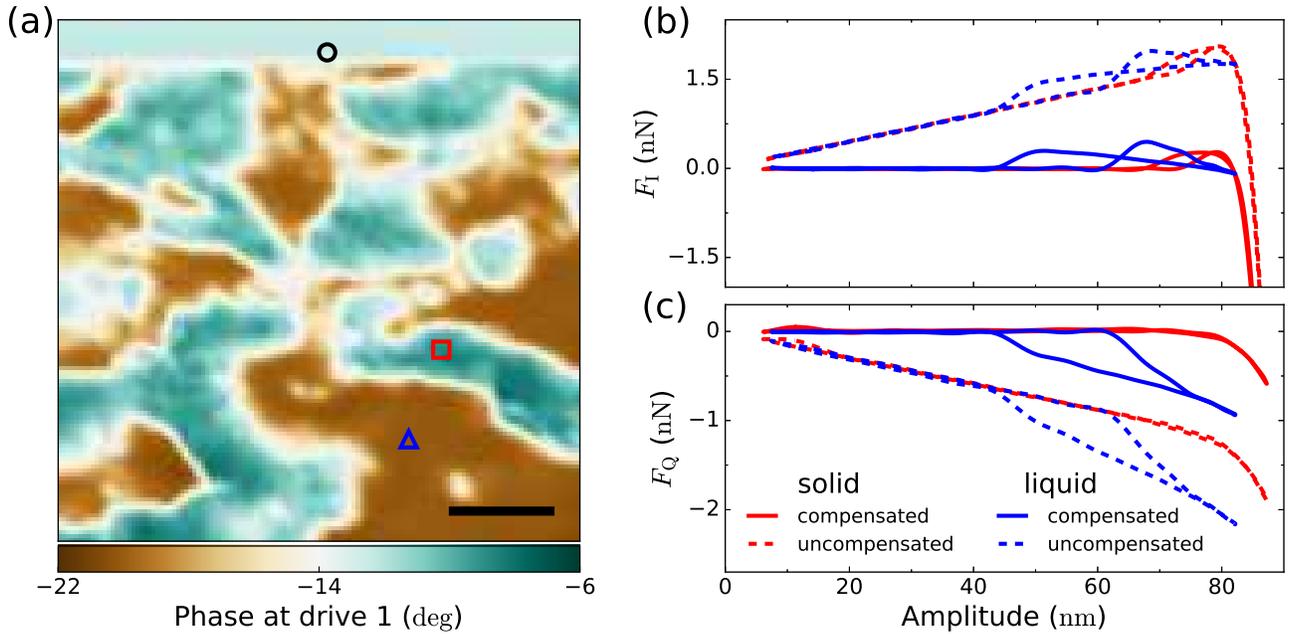}
    \caption{(a) Phase image at the first drive frequency on a silicone hydrogel (Young's modulus $0.35~\mathrm{MPa}$) in high humidity environment. The blue triangle and the red square mark the pixels for which the engaged spectrum is analyzed. The black circle marks the pixel where the lift motion was measured while scanning. At the top of the image the AFM was left scanning at the lift position, demonstrating the background forces are independent on the $xy$ position of the tip. The scale bar is $200~\mathrm{nm}$. (b, c) Dynamic force quadratures on a solid-like area (red) and on a liquid-like area (blue) at the pixels marked in the corresponding color. Dashed lines show uncompensated measurements and solid lines show compensation for background forces.}
    \label{fig:hydrogel}
\end{figure*}

We have shown how to mathematically treat the problem of compensating for arbitrary linear background forces, and we proposed a simple method to obtain their response function.
We now show an application of this method on soft material surfaces where background forces are typically rather large.

Figure~\ref{fig:PS_LDPE} shows dynamic force quadratures\cite{Platz2013B} on two areas of a polystyrene low-density polyethylene polymer blend (Bruker).
$F_\mathrm{I}$ is the force in phase with the cantilever motion integrated over one oscillation cycle, representing the conservative forces experienced by the cantilever at different oscillation amplitudes.
$F_\mathrm{Q}$ is the force quadrature to the cantilever motion integrated over one oscillation cycle, showing the dissipative interaction of the cantilever with its environment and with the surface.
The force quadratures represent a direct transformation of the measured data without any model assumptions, providing a physically intuitive way of analyzing the measured cantilever dynamics in terms of conservative and dissipative interactions.

At low amplitudes the uncompensated force quadratures (dashed lines in Fig.~\ref{fig:PS_LDPE}(b,c)) show a positive slope in $F_\mathrm{I}$, the signature of a long range attractive force.
The negative slope in $F_\mathrm{Q}$ is a signature of a linear damping, in addition to the damping contained in $\hat{\chi}$ which is calibrated far away from the surface.
Some hysteresis is also present in both sets of curves, indicating that the background forces are not purely of the type described by the special case of Eq.~(\ref{eq:polyfit_ho}).
Notably, the low amplitude background forces are the same for the two sets of curves, despite being measured over different areas of the sample with very different material properties, confirming that the background forces are not local to the AFM tip.

We used a drive scheme with $N_\mathrm{D} = 2$, measured $\hat{d}_\mathrm{lift}$ over a polystyrene area of the sample (black circle in Fig.~\ref{fig:PS_LDPE}(a)), then calculated $\hat{\chi}_\mathrm{BG}$ with $M = 1$ in Eq.~(\ref{eq:polyfit}) and applied its compensation to the measured data.
The solid lines in Fig.~\ref{fig:PS_LDPE}(b,c) show the force quadratures compensated for background forces.
The slope at low amplitudes is now missing, as well as most of the hysteretic effects (see inset of Fig.~\ref{fig:PS_LDPE}(b)).
We note that the compensation calculated over a polystyrene area of the sample has this effect not only for the force quadratures on the polystyrene, but also for those on polyethylene.
Taken together, these observations confirm the validity of the assumption that while the background forces change for every pixel of the image, their linear response function does not change during the whole scan.

Figure~\ref{fig:hydrogel} shows the same measurement and compensation procedure on a silicone hydrogel sample in a high humidity environment.
The sample presents alternating solid and liquid-like domains (Fig.~\ref{fig:hydrogel}(a)) with very different mechanical response as shown by the peculiar shapes of the force quadratures (Fig.~\ref{fig:hydrogel}(b,c)).
Also in this sample it is clear how the background force compensation corrects for the long-range attractive forces and increased dissipation, while at the same time preserving the interesting features of the tip-surface force and the peculiar hysteresis.
These will be discussed further in a forthcoming publication.

\begin{figure}
  \includegraphics[width=\columnwidth]{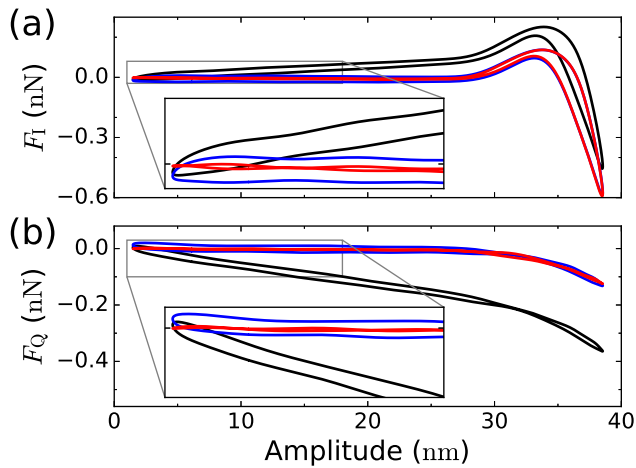}
  \caption{
  Comparison of dynamic force quadratures on polystyrene: uncompensated curves (black), compensated according to Eq.~(\ref{eq:polyfit}) (red), and according to Eq.~(\ref{eq:polyfit_ho}) (blue).
  The simplified case of Eq.~(\ref{eq:polyfit_ho}) fails to describe the hysteresis in the low amplitude region of the curves.
  }
  \label{fig:poly_vs_ho}
\end{figure}
For the dynamic force quadratures shown, the generic compensation according to Eq.~(\ref{eq:polyfit}) was used.
Figure~\ref{fig:poly_vs_ho} shows a comparison of these curves with the ones obtained from the special, simplified case of Eq.~(\ref{eq:polyfit_ho}).
While the simplified case can capture the general slope in the conservative and in the dissipative long range interactions, it fails to describe the hysteresis in the low amplitude part of the curves.
Thus, the background forces cannot be described by simply redefining the parameters of the cantilever resonance, indicating that a more thorough modeling of the interactions is required.

\begin{figure}
  \includegraphics[width=\columnwidth]{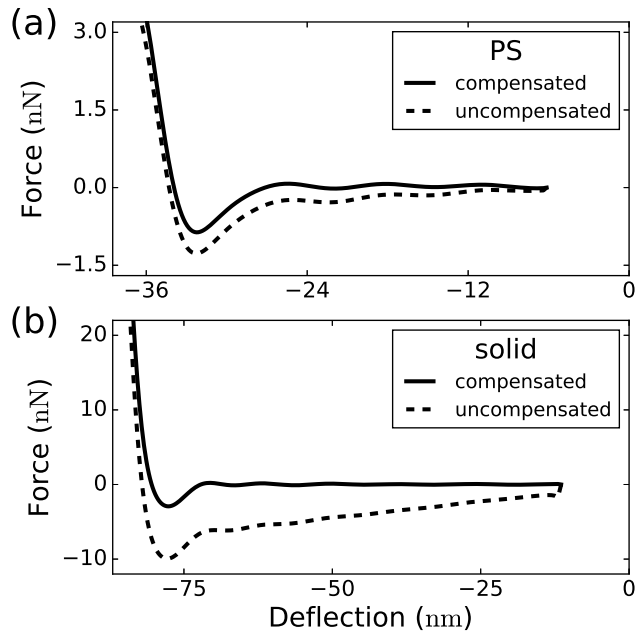}
  \caption{Tip-surface force on polystyrene (a) and a solid domain of hydrogel (b) reconstructed with amplitude-dependent force spectroscopy\cite{Platz2013A} (ADFS).
  The dashed lines show the uncompensated force, and the solid lines the force compensated for background interactions.}
  \label{fig:adfs}
\end{figure}
We have shown the effect of applying the background force compensation on the dynamic force quadratures $F_\mathrm{I}$ and $F_\mathrm{Q}$.
The compensation procedure is however general and can be applied to any force reconstruction method.
Once the frequency components of the compensated tip-surface force $\hat{F}_\mathrm{TS}$ are obtained from Eq.~(\ref{eq:correction}), $\hat{F}_\mathrm{TS}$ can be fed to any force reconstruction algorithm without further modifications.
As an example, in Fig.~\ref{fig:adfs} we calculate the tip-surface force on the polystyrene and on the solid domain of the hydrogel sample with amplitude-dependent force spectroscopy\cite{Platz2013A} (ADFS), with and without compensation for the background forces.
Not compensating for background forces would lead to overestimating the adhesion force by $47\%$ for polystyrene, and by $240\%$ for the solid domain of the hydrogel.
The observed peak-force would be instead underestimated by about $10\%$ in both cases (not shown in Fig.~\ref{fig:adfs}).

\section{Simulation of background forces}

Due to the nonlinear nature of the tip-surface interaction in dynamic AFM, the dynamics is typically simulated by numerically integrating the differential equation for the cantilever deflection $d(t)$
\begin{equation}
  \ddot{d} + \frac{\omega_0}{Q} \dot{d} + \omega_0^2 d = \frac{\omega_0^2}{k} \left( F_\mathrm{D} + F_\mathrm{TS} + F_\mathrm{BG} \right)
  \label{eq:simulation}
  .
\end{equation}
$F_\mathrm{D}(t)$ is a known function of time, whereas $F_\mathrm{TS}$ and $F_\mathrm{BG}$ are unknown.
A variety of models for $F_\mathrm{TS}$ are available\cite{VEDA} to simulate different types of both conservative and dissipative tip-surface interaction as function of the cantilever motion $d(t)$ and its velocity $\dot{d}(t)$, and even on the effective position of a moving surface model\cite{Haviland2016}.
On the other hand, no general model for simulating background forces is available, due to the different types of interaction that can give rise to this effect.

In the frequency domain, the background forces can be treated as an effective cantilever linear response function $\hat{\chi}^\prime$:
\begin{equation}
  \hat{F}_\mathrm{TS} = \hat{\chi}^{\prime-1} \hat{d}_\mathrm{eng} - \hat{F}_\mathrm{D}
  \label{eq:correction_chi}
  ,
\end{equation}
where
\begin{equation}
  \hat{\chi}^{\prime-1} = \hat{\chi}^{-1} - \hat{\chi}_\mathrm{BG}^{-1}
  \label{eq:chi_eff}
  .
\end{equation}
Transforming $\hat{\chi}^\prime$ into a differential equation is in general very difficult, however in the special case of Eq.~(\ref{eq:polyfit_ho}) it is possible to simply replace $\omega_0$ and $Q$ in Eq.~(\ref{eq:simulation}) with $\omega_0^\prime$ and $Q^\prime$ as defined by Eq.~(\ref{eq:w0_eff}) and Eq.~(\ref{eq:Q_eff}):
\begin{equation}
  \ddot{d} + \frac{\omega_0^\prime}{Q^\prime} \dot{d} + \omega_0^{\prime2} d = \frac{\omega_0^{\prime2}}{k} \left( F_\mathrm{D} + F_\mathrm{TS} \right)
  \label{eq:sim_chi_eff}
  .
\end{equation}

Alternatively, it is possible to more generally treat the effect of the background forces on the nonlinear response by introducing an effective drive force $\hat{F}^\prime_\mathrm{D}$:
\begin{equation}
  \hat{F}_\mathrm{TS} = \hat{\chi}^{-1} \hat{d}_\mathrm{eng} - \hat{F}^\prime_\mathrm{D}
  \label{eq:correction_F}
  ,
\end{equation}
where
\begin{equation}
  \hat{F}^\prime_\mathrm{D} = \hat{F}_\mathrm{D} + \hat{\chi}_\mathrm{BG}^{-1} \hat{d}_\mathrm{eng}
  \label{eq:F_eff}
  .
\end{equation}
Once the free, lift, and engaged motions are measured experimentally, Eq.~(\ref{eq:F_eff}) is used to determine the effective drive force $\hat{F}^\prime_\mathrm{D}(\omega)$ which is readily transformed to the time domain $F^\prime_\mathrm{D}(t)$ via the inverse Fourier transform.
The new differential equation
\begin{equation}
  \ddot{d} + \frac{\omega_0}{Q} \dot{d} + \omega_0^2 d = \frac{\omega_0^2}{k} \left( F^\prime_\mathrm{D} + F_\mathrm{TS} \right)
  \label{eq:sim_F_eff}
\end{equation}
can now be integrated numerically.

A comparison of the simulated motion $d_\mathrm{sim}$ using Eq.~(\ref{eq:sim_chi_eff}) or Eq.~(\ref{eq:sim_F_eff}), with the measured motion $d_\mathrm{eng}$, allows for numerical optimization to find the best-fit parameters of a nonlinear tip-surface force model.

\section{Conclusions}

We derived a mathematical procedure to account for long-range background forces in dynamic AFM, under the assumption of linear interaction and in the limit of a narrow band measurement.
Using intermodulation distortion to detect the onset of tip-surface forces, we accurately measure the background forces at the driven frequencies and extrapolate their linear response function to undriven frequencies.
Applying our procedure to experimental data we demonstrated compensation for background forces on dynamic force quadratures and ADFS force curves, measured on two different soft materials.
Given the generality of the compensation procedure and its ease of application to any type of dynamic force reconstruction, our method will be very useful for the determination of material properties with quantitative dynamic AFM.

\begin{acknowledgments}
The authors acknowledge financial support from the Swedish Research Council (VR), the Knut and Alice Wallenberg Foundation, and the Olle Engkvist Byggm\"astare Foundation.
Gunilla H\"agg (Star-Lens AB, \r{A}m\r{a}l, Sweden) is acknowledged for providing the silicone hydrogel sample.
\end{acknowledgments}

\appendix*

\section{Experimental and implementation details}
All measurements reported were performed in ambient atmosphere at room temperature with commercially available instrumentation.

The measurements on the polystyrene low-density polyethylene blend sample (Fig.~\ref{fig:imd}, Fig.~\ref{fig:PS_LDPE} and Fig.~\ref{fig:adfs}) were performed on a Bruker Dimension Icon Atomic Force Microscope.
The cantilever used was a MikroMasch HQ:NSC15/AlBS.
The fundamental flexural eigenmode of the cantilever and the detector were calibrated with the non-invasive thermal noise method\cite{Sader2012,Higgins2006}, which yielded resonance frequency $f_0=231.6~\mathrm{kHz}$, quality factor $Q=392.3$, mode stiffness $k=14.25~\mathrm{Nm^{-1}}$ and inverse optical lever responsivity $invOLR=88.63~\mathrm{nmV^{-1}}$.
The drive of the acoustic actuation was chosen to have a free oscillation amplitude of $50~\mathrm{nm}$ ($100~\mathrm{nm}$ peak-to-peak) for all the number of drives~$N_D$.
The scan in Fig.~\ref{fig:PS_LDPE} was performed with a drive composed of two frequencies separated by $500~\mathrm{Hz}$ and centered around the cantilever resonance.
The AFM feedback kept the amplitude at the lower drive frequency constant, at $80\%$ of its free value by adjusting the $z$-piezo extension.
The scan size was~$1.5~\mathrm{\mu m}$, the resolution 256x256 pixels and the pixel time~$2~\mathrm{ms}$, giving a total scan time of less than~$5~\mathrm{min}$.

The measurements on the silicone hydrogel (Fig.~\ref{fig:hydrogel} and Fig.~\ref{fig:adfs}) were performed on a JPK NanoWizard3 AFM.
The cantilever was a Bruker Tap525.
The thermal noise calibration yielded resonance frequency $f_0=470.3~\mathrm{kHz}$, quality factor $Q=626.0$, mode stiffness $k=86.39~\mathrm{Nm^{-1}}$ and inverse optical lever responsivity $invOLR=37.84~\mathrm{nmV^{-1}}$.
The amplitude feedback setpoint was $86\%$ of the free value.
The total free amplitude was~$105~\mathrm{nm}$ ($210~\mathrm{nm}$ peak-to-peak).
The scan size was~$1~\mathrm{\mu m}$, the resolution 128x128 pixels and the pixel time~$2~\mathrm{ms}$, giving a total scan time of about~$1~\mathrm{min}$.

For all the measurements reported, a multifrequency lock-in amplifier and dedicated control software (Intermodulation Products AB) were used as an add-on to the AFM hardware and software.

The automatic procedure for measuring the lift oscillation (see Sect.~\ref{sec:algorithm}) works as follows.
The scan size of the AFM is set to zero.
The scanning $z$-piezo is fully retracted (by setting a very high amplitude setpoint) and the free intermodulation distortion $\mathit{IMD}_\mathrm{free}$ is measured.
The original setpoint is restored to allow the tip to engage the surface again.
The setpoint is repeatedly increased in steps of~$0.1\%$ (causing the $z$-piezo to gradually retract) and the $\mathit{IMD}$ is measured for each of the setpoint values.
To allow the AFM feedback to stabilize and to lower the noise, at each step 100 pixels are acquired (typically requiring~$200~\mathrm{ms}$) and only the last 10 pixels are averaged and included in the calculation of $\mathit{IMD}$.
The procedure continues until the measured $\mathit{IMD}$ decreases to within~$3~\mathrm{dB}$ (a factor of~2) of $\mathit{IMD}_\mathrm{free}$.
The first measurement that satisfies $\mathit{IMD}<\mathit{IMD}_\mathrm{free} + 3~\mathrm{dB}$ defines the lift oscillation~$d_\mathrm{lift}$ which is then stored.
The original scan size and feedback setpoint are finally restored and the AFM scan is resumed.
The whole procedure requires less than one minute and is fully automated.

When the free and lift response is known, the background forces linear response function~$\hat{\chi}_\mathrm{BG}$ can then be calculated from Eq.~(\ref{eq:chi_bg}) at the driven frequencies (2 in the case of Fig.~\ref{fig:PS_LDPE} and Fig.~\ref{fig:hydrogel}, $N_\mathrm{D}$ in Fig.~\ref{fig:imd}).
To get~$\hat{\chi}_\mathrm{BG}$ at all the measured frequencies (32 in the case of Fig.~\ref{fig:PS_LDPE}, 42 in Fig.~\ref{fig:hydrogel}), a least square optimization routine (MINPACK) is used to fit Eq.~(\ref{eq:polyfit}) to the measured~$\hat{\chi}_\mathrm{BG}$.


%

\end{document}